\documentclass[prb,twocolumn,preprintnumbers,amsmath,amssymb,nofootinbib]{revtex4}

\usepackage{epsfig}
\usepackage{float}
\usepackage{graphicx}
\usepackage{dcolumn}
\usepackage{bm}
\usepackage{color,soul}
\usepackage[labelfont=bf,textfont=normalfont,singlelinecheck=off,justification=raggedright]{caption}
\usepackage[labelfont=bf,justification=justified]{subcaption}
\usepackage{caption}
\usepackage{fourier}
\usepackage[T1]{fontenc} 
\usepackage[english]{babel} 
\newcommand\addtag{\refstepcounter{equation}\tag{\theequation}}

\begin{document}

\title{Far-infrared Tamm  polaritons in a microcavity with incorporated graphene sheet}
\author{J. M. S. S. Silva,\textsuperscript{1} and M. I. Vasilevskiy\textsuperscript{1,2,*}}

\address{\textsuperscript{1}Physics Department and CFUM, University of Minho, 4710-057, Braga, Portugal,\\
	\textsuperscript{2}International Iberian Nanotechnology Laboratory, 4715-330, Braga, Portugal\\
\textsuperscript{*}mikhail@fisica.uminho.pt
}
%

\begin{abstract}
Tamm polaritons (TPs) are formed at the interface between two semi-infinite periodic dielectric structures (Bragg mirrors) or other reflectors. 
Contrary to usual surface polaritons, TPs exist inside the "light cone", 
even though their amplitude also decreases exponentially with the distance from the interface 
as it is characteristic of evanescent waves. 
They couple to elementary excitations in the materials or structures that form the interface, such as metal plasmons or semiconductor excitons. 
Here we discuss the formation of TPs in the far-infrared (FIR) spectral range, in the optical-phonon reststrahlen band of a polar semiconductor such as GaAs, with a Bragg reflector (BR) as the second mirror. 
Their dispersion relation and the frequency window for the TP existence are discussed for a GaAs-BR interface. 
Structures containing a gap between the two reflectors are also considered. 
Further investigation is performed on a structure containing a layer of graphene between the two reflectors.
\end{abstract}
\maketitle
{\it \textbf{Introduction.}} 
Confinement of light near an interface between two materials provides the framework for its manipulation at nanoscale and usually is achieved by using metallic materials and nanostructures  \cite{Shalaev:10,Gramotnev:10}, metamaterials \cite{Dickson:15}, and metal-like optical properties of excitonic \cite{Sara_2016} or phononic \cite{Straude:16} materials.  In these structures, the electromagnetic (EM) field confinement is caused by the coupling to elementary excitations in the material, such as plasmons, excitons or phonons. 
The possibility of the existence of interface EM waves at the interface between two reflecting media, 
analogous to electronic Tamm states arising at the surface of a crystal because of the broken translational symmetry \cite{Tamm_vol1}
has been predicted theoretically for two semi-infinite periodic dielectric structures (that can be named superlattices or 1D photonic crystals or simply Bragg reflectors, BRs) \cite{KavokinOTSs}, later for a gold slab combined with a dielectric BR  \cite{TammPlasmonTP}, in a periodic metal/dielectric structure  \cite{Zhou:10} and, more recently, for a metamaterial composed of a periodic sequence of metal or graphene sheets intercalated by dielectric layers \cite{Iorch_2011,Smirnova_2014}.
Such waves were called Tamm polaritons (TPs)  or optical Tamm states (OTS).
Unlike the electronic Tamm states, OTS cannot occur at the  
free surface of a photonic crystal but exist at the interface between two photonic 
structures having overlapping photonic band gaps (also called stop bands). 
Contrary to the usual surface polaritons, TPs appear inside the "light cone", 
even though their amplitude also decreases exponentially with the distance from the interface 
as it is characteristic of evanescent waves.\\
%
\begin{figure*}[t]
	\centering
	\includegraphics[scale=0.4]{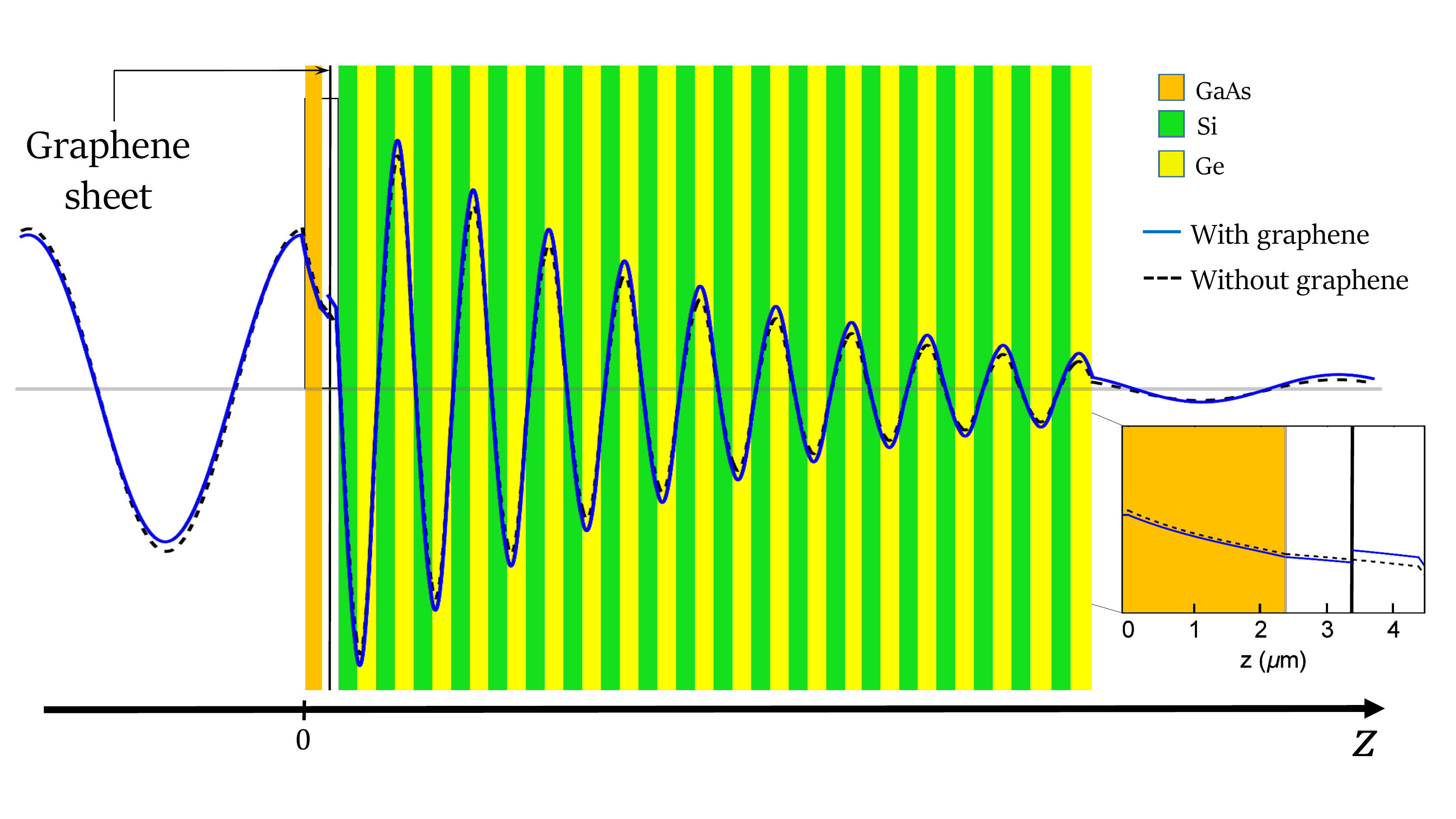}
	\caption{Schematics of the studied heterostructure and calculated profile of the magnetic field amplitude corresponding to OTS and external EM wave in air used for probing this confined state considering a structure with and without a graphene sheet (blue solid line and black dashed line, respectively), for a frequency of $35.60$ meV and $35.48$ meV, respectively, corresponding to the minimum of reflectance in each case. Notice the discontinuity of the magnetic field profile across the graphene layer introduced in the middle of the gap (assuming graphene's Fermi energy $=$ $0.5$ eV) and the continuity of the field profile in the absence of the graphene layer.}
	\label{fig:scheme_structure}
\end{figure*} 
The existence of Tamm plasmon-polaritons has been demonstrated experimentally for GaAs/AlGaAs superlattices covered with a gold layer \cite{sasin_2008, sasin_2010}.
More recently, coupling of OTS to excitons
in a layer of aggregated dye molecules \cite{Sara_2016} and in a 2D semiconductor layer \cite{lundt_2016} has been demonstrated,
leading to the formation of room-temperature exciton-polaritons, mixed excitations  
where the excitons are localized in a very
thin layer while the EM field is confined (at a much larger scale) within a planar microcavity \cite{mikhail_2015}. 
Exciton-polaritons have been a popular research topic in the past years and a number of polariton confinement techniques have been developed in the visible range of the EM spectrum \cite{schneider_2016}. %
While the formation of plasmon-polaritons,
with an exponentially decaying amplitude inside the metal, is due to the negative dielectric constant of the metal below its plasma frequency, in the latter case the exciton has no role in the formation of the OTS  
(formed by non-excitonic materials)
but just couples to it.
The metal film was used as adjustable means for making a microcavity (sometimes called Tamm microcavity) \cite{sebald_2015}.\\
The observation of hybrid exciton-polaritons in the regime of strong coupling between large-radius Wannier type GaAs excitons, tightly bound excitons in a MoS$\text{e}_2$ monolayer, and cavity photons within a Tamm-plasmon-polariton device paves the way towards a manifold of applications of hybrid exciton-polaritons \cite{lundt_2017}. 
In particular, the formation of a condensate of exciton-polaritons in such a structure was observed \cite{lundt_2018}, contributing to the investigation of highly efficient, ultra-compact polariton-based light sources and valleytronic devices.\\
Closer to immediate practical applications, a recent investigation of a hybrid Tamm-plasmon-polariton sensor for blood components detection was performed \cite{maji2018}.    
The study of advanced light trapping schemes using optical Tamm states in organic solar cells, 
which are composed of photonic crystal bilayers with high refractive index contrast stacked outside the organic layers has shown enhanced photon absorption \cite{zhang_2013}.
The discovery of a tunable optical Tamm state at the interface between a photonic crystal and a film with non-uniformly varying refractive index may be appealing to applications such as optical interference filters and environmental sensing devices \cite{zheng_2018}.\\
The purpose of the present work is to extend these ideas to a different (namely, far-infrared (FIR)) spectral range and to describe theoretically Tamm polariton states that can form at the interface between a polar semiconductor (such as GaAs) and a Bragg reflector suitable for the FIR range.
The semiconductor acts as a phononic mirror  
within the frequency range between the transverse ($\omega_{\text{TO}}$) and longitudinal ($\omega_{\text{LO}}$) optical phonon frequencies (called reststrahlen band), where
the real part of the dielectric function of the semiconductor is negative \cite{Yu-Cardona}. 
The Tamm polaritons are expected in the frequency range where the BR stop band and the reststrahlen band of GaAs overlap. 
We shall explore their dispersion relation considering both the "pure" Tamm state $\left( \delta = 0\right)$ and cavity modes $\left( \delta \neq 0\right)$. 
In addition, we shall consider the effect of insertion of a 2D conductor (graphene) layer into the cavity.
The studied structure is presented in Fig. \ref{fig:scheme_structure}.\\
Graphene, a two-dimensional form of carbon, possesses electronic, mechanical and optical properties \cite{geim2009,neto_guinea_peres_novoselov_geim_2009}. The high frequency conductivity and, consequently, the optical transparency of graphene can be controlled by changing its Fermi level by means of gating \cite{li_2008}, making it a transparent conductor with tunable conductivity. 
Doped graphene supports $p-$polarized surface plasmon-polaritons in the THz range and it has given rise to a broad field of research, both theoretical \cite{bludov_2013} and experimental \cite{Luo_2013}, including proposals of graphene-based tunable metamaterials targeting different properties \cite{Smirnova_2014,Ju_2011,Xiang_2014,Bludov:15}.  
Recently, it has been shown \cite{Wang:17}  that an enhanced absorption of THz radiation can be achieved in a composite structure where graphene is deposited on a BR, separated by a dielectric. It can happen due to the excitation of TPs at the graphene covered surface, auxiliated by the graphene sheet.
Here we consider a different structure, a Tamm-type microcavity formed by a phononic mirror and a Bragg reflector, and also with a different idea.
Inserting a graphene layer between the two mirrors may result in the coupling of the graphene's plasmon to the FIR Tamm polaritons, thus introducing a control mechanism over cavity modes. This possibility is explored in the present work. \\

%
%
{\textit{\textbf{OTS dispersion relations.}}} Firstly, we will consider a structure without graphene and then we will investigate a structure containing a graphene sheet.\\
Let us consider two heterostructures separated by a gap of thickness $\delta $. One of them is a Bragg reflector consisting of $N$ "unit cells" composed of a layer of material A (e.g. Si) and a layer of a material B (e.g. Ge). For simplicity, we shall assume that these layers have the same thickness, $d$.
The transfer matrix approach \cite{novotny_nanooptics} is very convenient for the study of the reflectance and transmittance of such a structure, which are defined by the Fresnel coefficients. 
We shall consider $p-$polarized EM fields and use transfer matrices, $\hat{T}$, defined in the basis of the magnetic field $H$ (directed perpendicular to the plane of Fig. \ref{fig:scheme_structure}) and the electric field component $E_x$. For instance, for the first layer the BR we have:
\[
\left(
\begin{tabular}{ccc}
$H_{A} (\mathbf{r},t)$  \\
$E_{Ax}(\mathbf{r},t)$  
\end{tabular}
\right)_{z={d^{-}}}
=
\hat{T}_{A} \cdot
\left(
\begin{tabular}{ccc}
$H_{A} (\mathbf{r},t)$  \\
$E_{Ax}(\mathbf{r},t)$ 
\end{tabular}
\right)_{z={0^{+}}} \addtag \text{,}
\]
where $\hat{T}_{A}$ denotes the transfer matrix of the medium A. 
For a multilayer structure, the transmission and reflection Fresnel coefficients, $\hat{r}_p$ and $\hat{t}_p$ respectively, can be obtained through the following relation \cite{icton2018}: 
\[ 
\left(
\begin{tabular}{ccc}
$1 + \hat{r}_p$  \\
$\frac{c k_{1z}}{\epsilon_1 \omega}(1-\hat{r}_p)$   
\end{tabular}
\right)
=
\hat{T}_N^{-1} \cdot
\left(
\begin{tabular}{ccc}
$\hat{t}_p$  \\
$- \frac{c k_{3z}}{\epsilon_3 \omega}\hat{t}_p$   
\end{tabular}
\right) \text{.}  \addtag \label{TM_method3N1}
\]
Here $k_{iz}$ is the $z$ component of the wavevector in the medium on the left ($i=1$) and on the ($i=3$) right of the multilayer structure,
\begin {equation}
k_{iz}=\sqrt{\epsilon_i  \left( \frac{\omega}{c} \right) ^2 - q^2  }\; ,
\label{k_z}
\end {equation} 
$\epsilon_i$ is the corresponding dielectric constant, $\omega $ and $q$ denote the frequency and transverse wavevector component, respectively, and
$\hat{T}_N^{-1}$ is the inverse transfer matrix of the multilayer structure, e.g. for the $N$-period BR:
\begin{align}
\hat{T}_N^{-1} \equiv \left( \hat{T}_{A}^{-1}\cdot \hat{T}_B^{-1} \right)^N \text{,}
\label{eq:TM_N}
\end{align}
where $\hat{T}_B$ is the transfer matrix of layer B.
The explicit form of the transfer matrices $\hat{T}_A$ and $\hat{T}_B$  is given in the Appendix.
We assume that $\epsilon_A$ and $\epsilon_B$ entering these relations are real constants. In the derivation of the OTS dispersion relation, BR will be considered as heterostructure 2  and its transfer matrix and Fresnel coefficient will be denoted $\hat T_2(= \hat T_N)$ and $ \hat{r}_2^{(p)} $, respectively.    
\par
\begin{figure}
	\centering
	\includegraphics[scale=0.3]{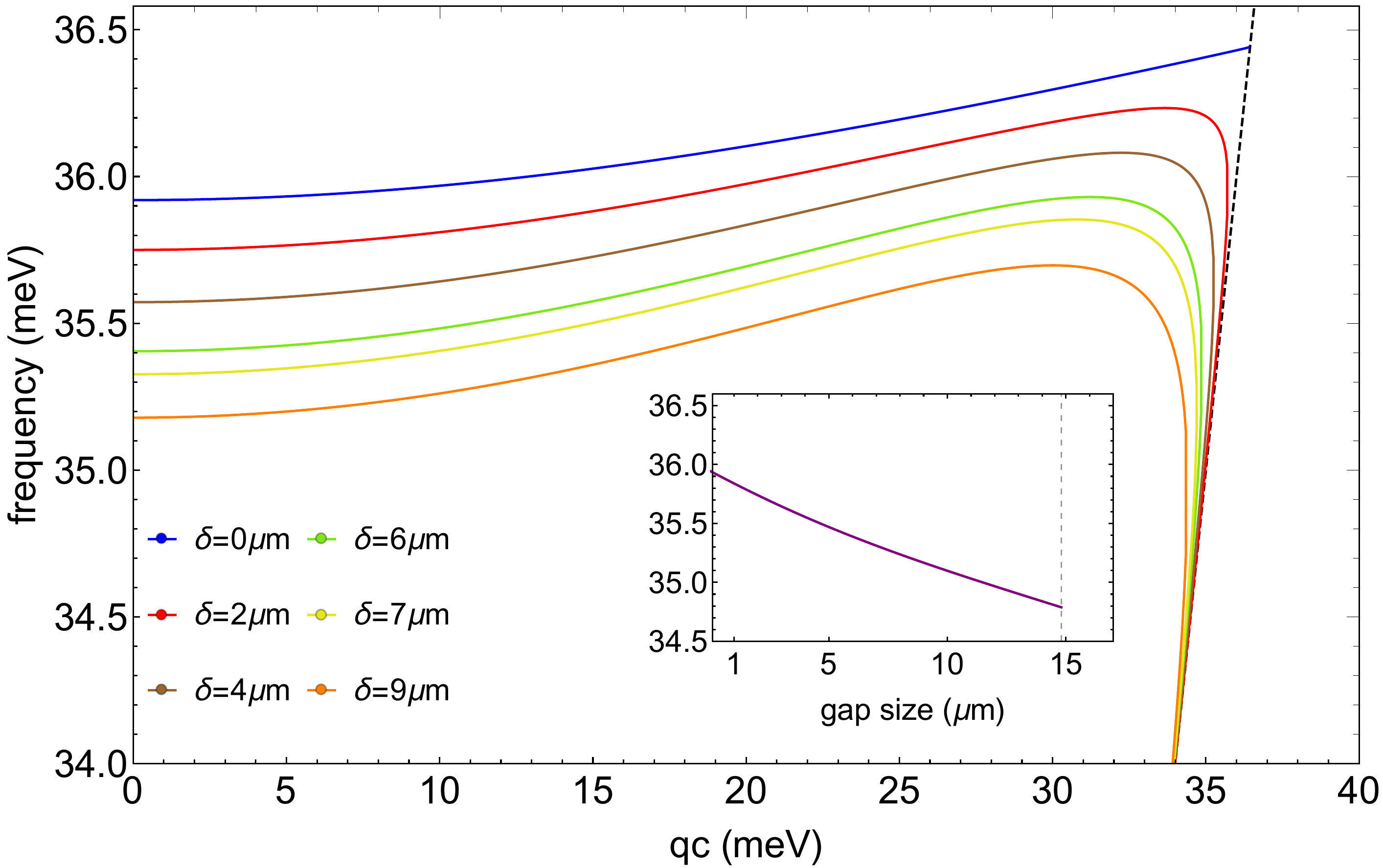}
	\caption{Eigenmode frequencies, solutions of the dispersion equation (\ref{eq:phase_match})  
		calculated for different gap values, for the structure without graphene. The Si/Ge layer thickness is $2.4\,\mu \text{m}$ and the GaAs slab thickness is $10 \mu \text{m}$, the GaAs phonon damping was neglected. Both the frequency and the wavevector values are presented in energy units. The inset shows the variation of the $q=0$ mode with the gap size.}
	\label{fig:disp_rel_delta}
\end{figure}

The other heterostructure in our case is just a homogeneous polar semiconductor (let us call it GaAs for definiteness). Its dielectric function owing to the polar optical phonon response is given by \cite {Yu-Cardona}: 
\begin{equation}  
\label{DF}
\epsilon_{GaAs} (\omega )=\epsilon _{\infty}\left (1+ \frac {\omega _{\text {LO}}^2-\omega
	_{\text {TO}}^2} {\omega _{\text {TO}}^2-\omega ^2-i\omega \Gamma_{\text {TO}}}\right ),
\end{equation}
where $\Gamma_{\text {TO}}$ is a phonon damping parameter and  $\epsilon _{\infty}$ is the high frequency dielectric constant.
The tranfer matrix  $(\hat T_1)$ is related to the dielectric function and thickness of the slab by the same equation (\ref  {TM_AB}) as for A and B layers of the Bragg reflector and the Fresnel reflecttion coefficient  (to be denoted $ \hat{r}_1^{(p)} $) is obtained from equation equivalent to Eq. (\ref {TM_method3N1}).\\ 
Now we shall consider the whole structure of Fig. \ref{fig:scheme_structure} consisting of the GaAs slab, the gap of thickness $\delta$ and the BR and assume that there is no incident wave but there are only "transmitted" (i.e. outgoing) waves at both sides of the whole structure.
Using the field matching condition at $z=0$, we arrive at the following equation:
\begin{equation}
\label{eq:TAMM_DISP1}
\hat{r}^{(p)}_1 \hat{r}^{(p)}_2 e^{2 i k \delta }=1 \text{ ,}
\end{equation}
where $\hat{r}^{(p)}_1$ and $\hat{r}^{(p)}_2$ correspond to the Fresnel reflection coefficients of medium 1 and medium 2, respectively.
In the case $\delta \rightarrow 0$ (two heterostructures back-to-back), it reduces to $\hat{r}^{(p)}_1 \hat{r}^{(p)}_2 =1 \text{ ,}$ which has been presented in Ref. \cite{TammPlasmonTP}. 
We emphasize that the Fresnel coefficients with labels 1 and 2 in Eq. (\ref{eq:TAMM_DISP1}) correspond to each heterostructure alone, in vacuum. Equation (\ref{eq:TAMM_DISP1})  requires that 
\begin{equation}
\label{eq:mod_match}
\vert \hat{r}^{(p)}_1 \vert = \vert \hat{r}^{(p)}_2 \vert = 1 \text{ ,}
\end{equation}
and
\begin{equation}
\label{eq:phase_match}
\Delta \phi = \phi_1 + \phi_2 + 2k\delta = 2 \pi \text{m} \text{ ,}
\end{equation}
where $\phi_{1,2}=\text{arg}\left( \hat{r}^{(p)}_{1,2} \right)$ and m is an integer. Assuming that Eq.  (\ref{eq:mod_match})  is (approximately) satisfied,
the phase matching condition (\ref{eq:phase_match}) determines the Tamm mode, i.e. it is an implicit dispersion relation, $\omega (q)$.\\
The materials chosen for our model calculations are GaAs, a semiconductor with $\omega_{{\text{TO}}} = 268\, {\text{cm}}^{-1}$ and $\omega_{\text{LO}} = 292\, {\text{cm}}^{-1}$ \space (Ref. \cite{Yu-Cardona}), and silicon and germanium for the BR, which are two compatible materials,  non-absorbing in the relevant frequency range.
The reflection coefficient of the Bragg reflector composed of Si and Ge with $d=2.4\, \mu$m shows a well defined stop band of width $\approx 4\, $meV centred at $\approx 35\, $meV, which covers most of the reststrahlen band of GaAs and $\vert \hat{r}^{(p)}_2 \vert \approx 1$ with a fairly high precision already for $N \sim 20$.  
The Fresnel reflection coefficient for a thick GaAs slab in air is  $ \hat{r}^{(p)}_1\approx (\sqrt  {\epsilon_{GaAs}}-1)/(\sqrt  {\epsilon_{GaAs}}+1)$. Its modulus nearly equals to unity for $\omega_{{\text{TO}}} < \omega < \omega_{{\text{LO}}} $ if we neglect the phonon damping, so it 
acts as a good mirror in this frequency range, similar to a metal but less lossy because the imaginary part peaks at $\omega_{\text{TO}}$ and the width of this peak is just few cm$^{-1}$ for a good crystal, while the width of the reststrahlen band is typically a few tens of cm$^{-1}$.
As known, because of this multilayer structures containing polar semiconductors support evanescent waves named surface phonon-polaritons, which occur in their reststrahlen band\cite{vinogradov_2002}.\\
The Tamm polariton dispersion curves obtained from the phase matching  equation (\ref{eq:mod_match}) for different values of the cavity thickness are presented in Fig. \ref{fig:disp_rel_delta}. 
They are approximately parabolic near $q = 0$, so we can say that the polaritons have some effective mass. For $\delta \neq 0$, the dispersion curves shift downwards within the reststrahlen band and their shape deviates from parabolic and becomes non-monotonic within the light cone. There is a critical size of the microcavity, above which the mode ceases to exist, $\approx 15\,\mu$m for $q=0$ (see inset in Fig. \ref{fig:disp_rel_delta}). 
%

Let us now introduce a graphene sheet inside the gap as shown in Fig. \ref{fig:scheme_structure}. 
%
%
%
\begin{figure*}[t]
	\captionsetup[subfigure]{justification=centering}
	\centering
	\begin{subfigure}{0.49\textwidth}
		\includegraphics[scale=0.47]{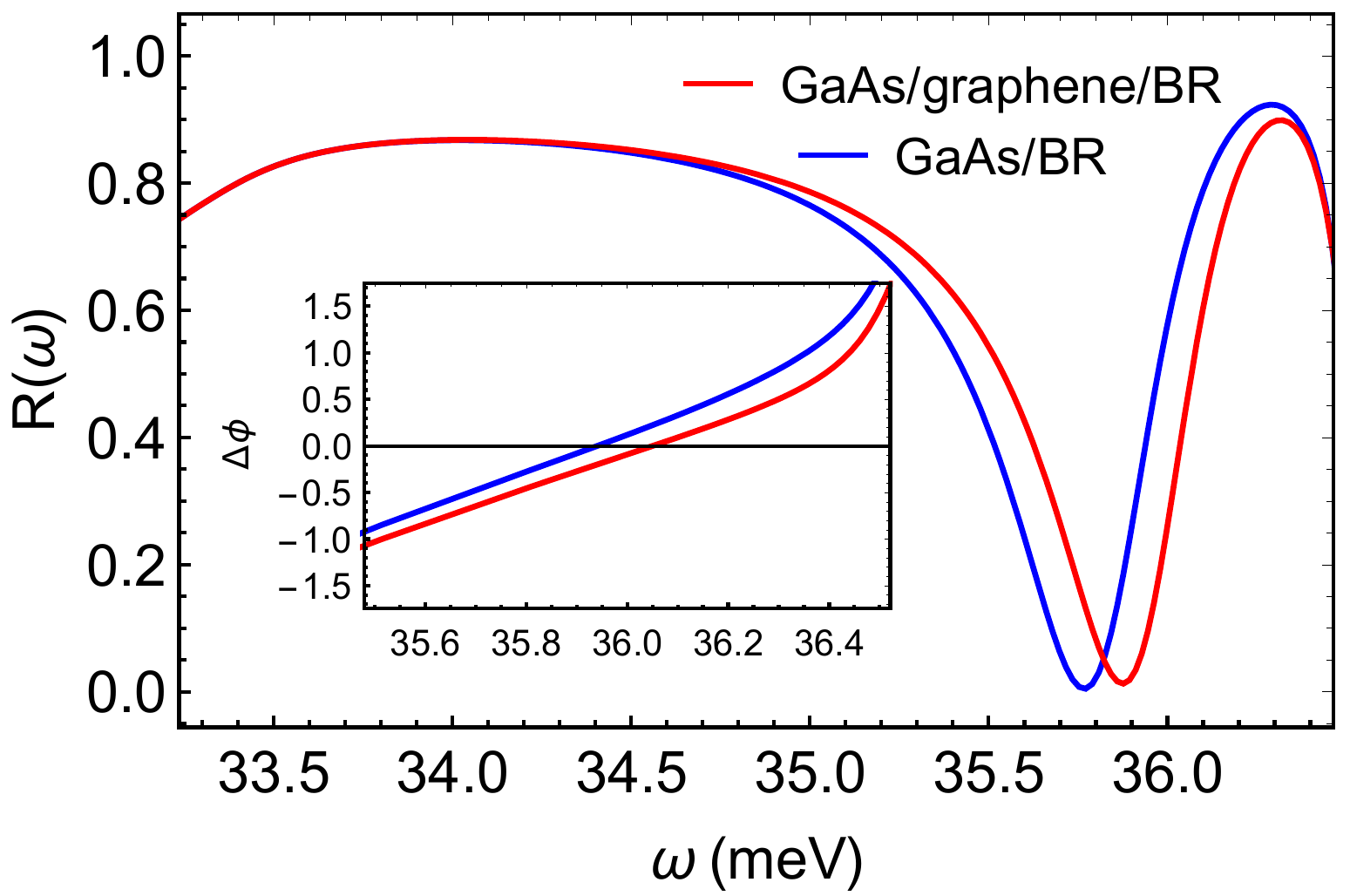}
		\caption{}
		\label{fig:reflec_graph_nograph}
	\end{subfigure}
	\begin{subfigure}{0.49\textwidth}
		\includegraphics[scale=0.47]{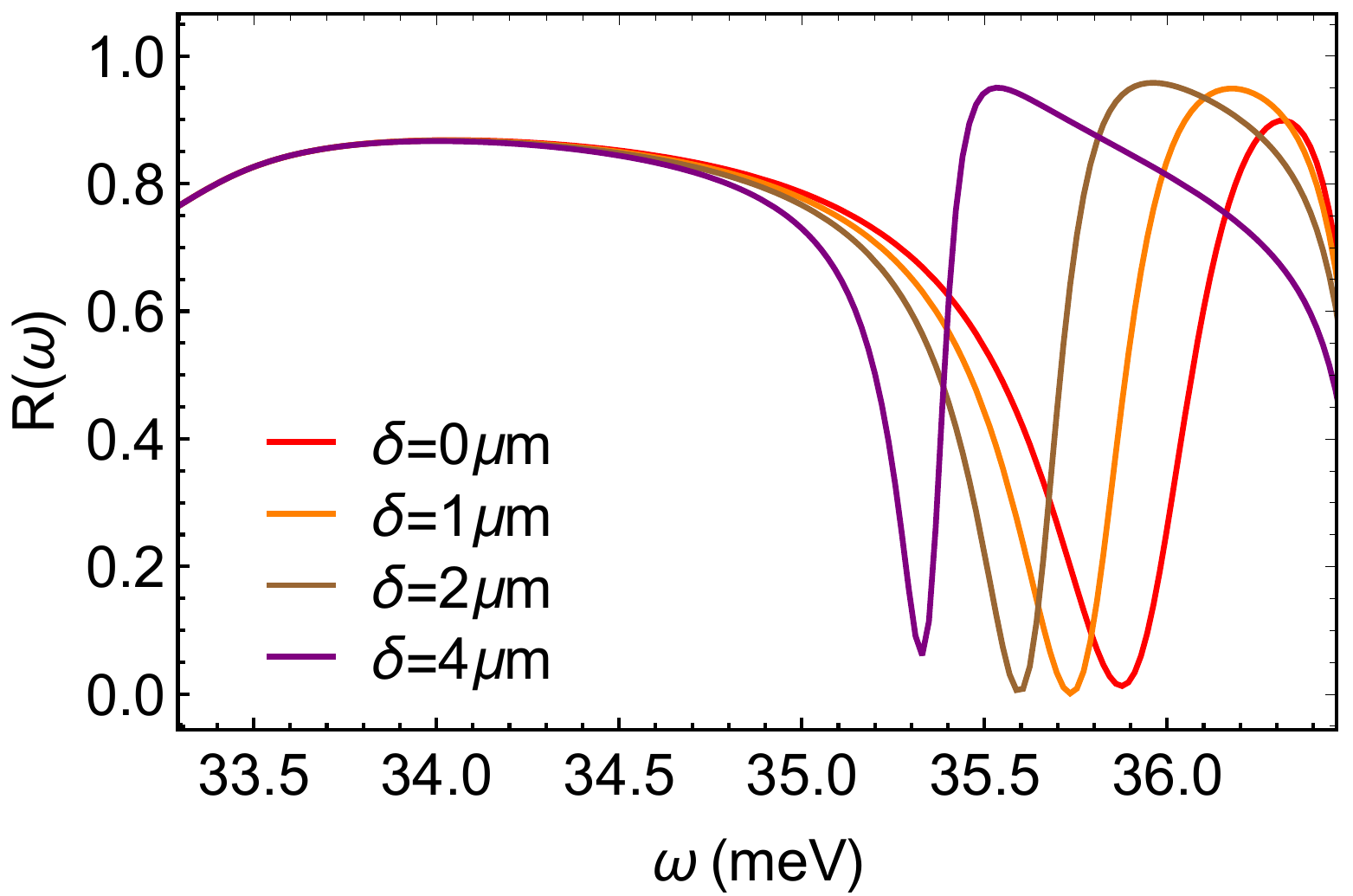}
		\caption{}
		\label{fig:reflec_graph_diffgaps}
	\end{subfigure}
	\caption{ Normal incidence reflectance spectra of {(a)} the structure without gap ($\delta=0$), with and without graphene layer at the GaAs/BR interface, and {(b)}  structures with different gap widths containing a graphene layer placed at the center of the microcavity. Parameters: GaAs layer thickness $2.37 \, \mu$m, $d=2.4 \, \mu$m, $N=$30 and $E_F=0.5\,$eV.}
	\label{fig:graphene_sheet1}
\end{figure*}
Following the same procedure as before, one can derive the dispersion relation from the following matching condition:
\[
\hat{T}_{\frac{\delta}{2}} \cdot \hat{T}_{Gr} \cdot \hat{T}_{\frac{\delta}{2}} \cdot
\left(
\begin{tabular}{ccc}
$1+\hat{r}^{(p)}_2$\\
$\frac{c k}{\omega}(1- \hat{r}^{(p)}_2 )$
\end{tabular}
\right)
=
D
\left(
\begin{tabular}{ccc}
$1 + \hat{r}^{(p)}_1$\\
$\frac{c k }{\omega} (\hat{r}^{(p)}_1 -1)$
\end{tabular}
\right) \text{,}\addtag \label{eq:tamm_pol_graphene11}
\] 
where $D$ is a constant, $k$ denotes the $z$ component of the wavevector in the gap, $\hat{T}_{\frac{\delta}{2}}=\exp {(ik\frac{\delta}{2})}\hat{I}$ ($\hat{I}$ is the unit matrix) and $\hat{T}_{Gr}$ is the graphene's transfer matrix, which is obtained by using the discontinuity condition for the transverse magnetic field across a 2D conductor \cite{bludov_2013}:
\[ 
\hat{T}_{Gr}
=
\left(
\begin{tabular}{ccc}
1 & $\frac{4 \pi}{c} \sigma \left( \omega \right)$  \\
0 & 1
\end{tabular}
\right) \text{ ,} \addtag \label{eq:TM_graphene}
\]
with $\sigma \left( \omega \right)$ being the graphene's 2D optical conductivity given by the Drude-type relation \cite{neto_guinea_peres_novoselov_geim_2009,bludov_2013}: 
\begin{equation}
\sigma \left( \omega \right) =  \sigma_0 \frac{4 E_F}{\pi} \frac{1}{\Gamma - i \hbar \omega} \text{ .}
\end{equation}
Here $\sigma_0=\pi e^2 / \left( 2 h \right)$, $e$ is the electron charge, $E_F>0$ denotes graphene's Fermi energy and $\Gamma$ is a damping parameter determined by electron scattering.
From Eq. (\ref{eq:tamm_pol_graphene11}), by eliminating $D$ we have: 
\begin{equation}
\label{eq:TAMM_DISP_graphene11}
\resizebox{0.43\textwidth}{!}{$\hat{r}^{(p)}_1 \hat{r}^{(p)}_2 e^{i k \delta } \left [\frac{2 \pi \sigma \left( \omega \right) k}{\omega} -1 \right ] - \frac{2 \pi \sigma \left( \omega \right) k}{\omega } \left[\hat{r}^{(p)}_1 +
\hat{r}^{(p)}_2 \right] + e^{-ik\delta}\left [ \frac{2 \pi \sigma \left( \omega \right) k}{\omega} + 1 \right ] = 0 \text{ .}
$}
\end{equation}
We can rewrite Eq. (\ref{eq:TAMM_DISP_graphene11}) as:
\begin{equation}
\label{eq:TAMM_DISP_graphene2}
\hat{r}^{(p)}_1 \hat{r}^{(p)}_2 e^{2i k \delta } =1 + \Delta\left(k,\omega\right) \text{ ,}
\end{equation}
where
\begin{equation}
\label{eq:TAMM_DISP_graphene3}
\Delta\left(k,\omega\right)=\frac{2 \pi \sigma \left( \omega \right) k}{\omega} \left(\hat{r}^{(p)}_1 e^{i k\delta} - 1 \right)\left(\hat{r}^{(p)}_2 e^{i k\delta} - 1 \right) \text{ .}
\end{equation}
By considering $\vert \hat{r}^{(p)}_1 \vert \simeq \vert \hat{r}^{(p)}_2 \vert \simeq 1 $ and approximating $\text {Re}\Delta \left(k,\omega \right) \ll 1$, one can write the following phase matching condition in the presence of graphene:
\begin{equation}
\label{eq:TAMM_DISP_graphene4}
\phi_1 + \phi_2 + 2 k \delta =\text{arg}\left[ 1 + \Delta\left( k,\omega \right) \right] \text{ .}
\end{equation}
We may say that the insertion of graphene into the Tamm microcavity introduces an additional phase shift, $\approx \text {arctg} [\text {Im }\Delta\left( k,\omega \right)] $, proportional to the graphene conductivity. As we shall see in the next 
section, it leads to a shift of the OTS frequency, which depends on the graphene's Fermi level.\\

\begin{figure}[H]
	\captionsetup[subfigure]{justification=centering}
	\centering
	\begin{subfigure}{0.49\textwidth}
		\includegraphics[scale=0.53]{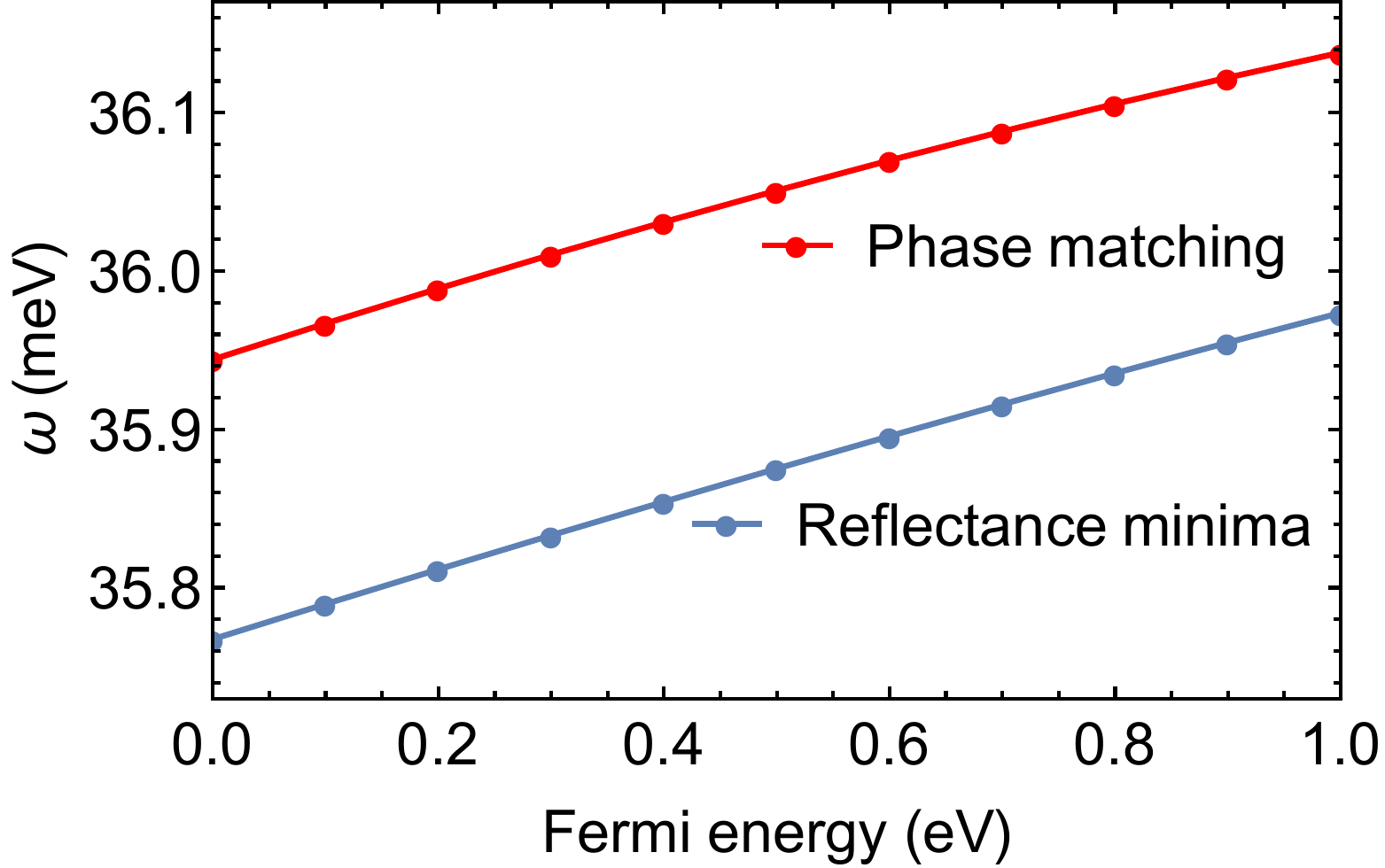}
		\caption{}
		\label{fig:fermi_level_withgrap}
	\end{subfigure}
	\begin{subfigure}{0.49\textwidth}
		\includegraphics[scale=0.27]{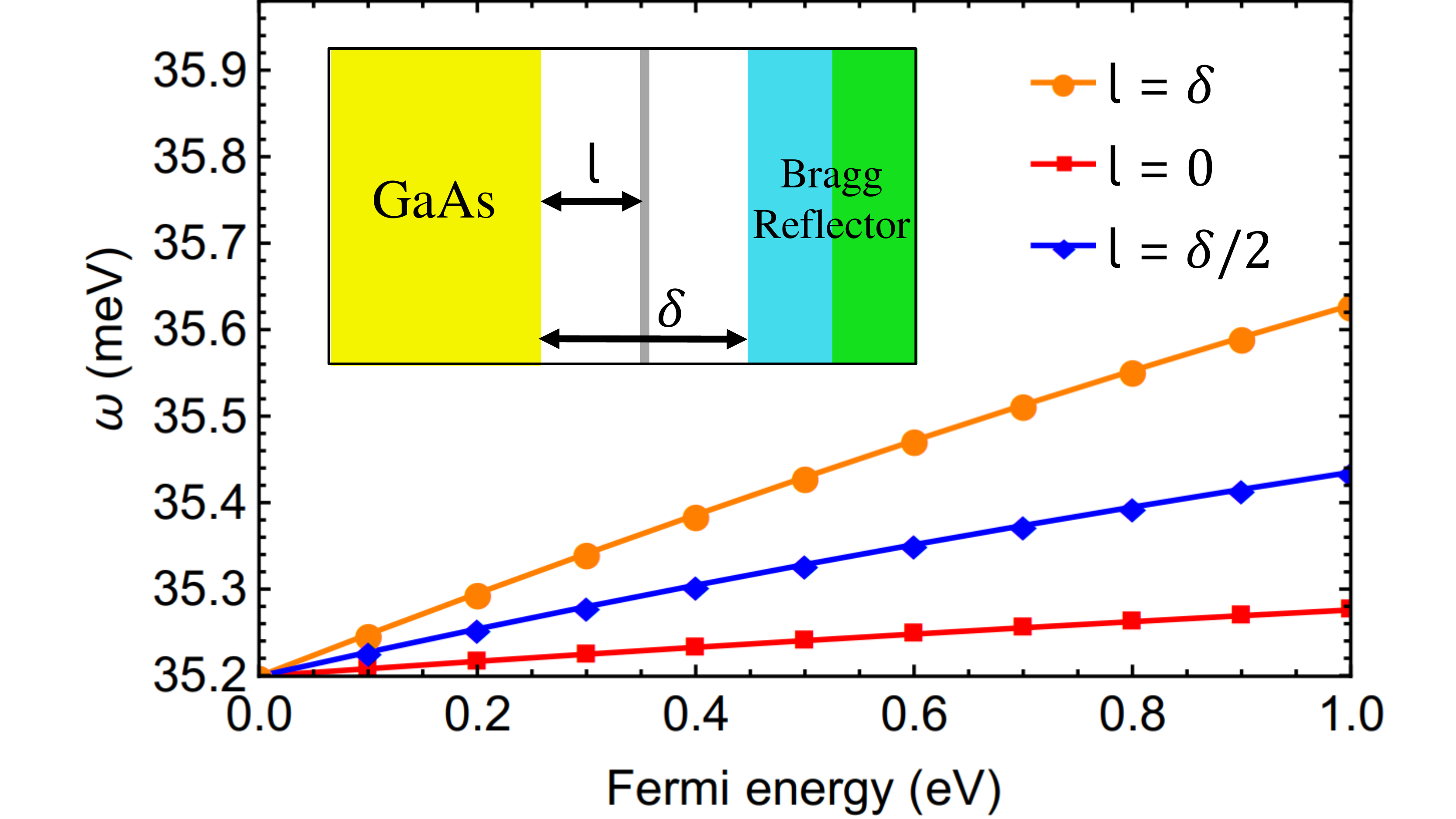}
		\caption{}
		\label{fig:fermi_level_diff_graphene_positions}
	\end{subfigure}
	\caption{Dependence of the OTS  frequency upon graphene's Fermi energy: {(a)} extracted  from normal incidence reflectance spectra of GaAs/graphene/BR structure and obtained by the phase matching condition (\ref{eq:TAMM_DISP_graphene4}); {(b)} for different locations of the graphene sheet as shown in the inset (the values are extracted from the reflectance spectra). Parameters are the same as in Fig. \ref{fig:graphene_sheet1}.}
	\label{fig:graphene_sheet2}
\end{figure}

%
{\it \textbf{Probing the Tamm states.}}
How can the OTS be observed? One may do it by shining far-infrared ( FIR) radiation onto the GaAs outer surface and measuring the reflectivity spectrum of the whole structure. 
Coupling of the incident (propagating) wave to the Tamm mode occurs owing to the (small) overlap between two decaying waves, one originated by the incident EM wave and the other corresponding to the Tamm state as depicted in Fig. \ref{fig:scheme_structure}.
If the overlap becomes large, the OTS will be strongly influenced by the incident wave and its frequency will be different from that of the proper mode, so the thickness of the GaAs slab (one of the "cavity walls") cannot be too small.
On the other hand, if its thickness is too large, the coupling will be too weak to measure. 
The reflectivity spectrum for the structure without graphene, for the GaAs slab thickness $2.37 \, \mu \text{m}$ is shown in Fig. \ref{fig:graphene_sheet1}. It shows a sharp dip at frequency close to that corresponding to $\Delta \phi=0$.
Even though GaAs slab is not a perfect mirror for the frequencies close to $\omega_{LO}$ (since the real part of the dielectric function is only slightly negative and $\vert \hat{r}^{(p)}_{GaAs} \vert < 1$), the phase matching condition (\ref{eq:phase_match}) is a fairly good indication of the OTS as illustrated in Fig. \ref{fig:graphene_sheet1}. 
The insertion of a graphene sheet into the cavity results in a shift of the reflectance minimum corresponding to the OTS mode towards a higher frequency (see Fig. \ref{fig:graphene_sheet1}(a)).
This shift is comparable to that due to the variation of the microcavity width, $\delta $ (see Fig. \ref{fig:graphene_sheet1}-b).  
The graphene-induced OTS shift increases with the increase of the Fermi energy (see Fig. \ref{fig:graphene_sheet2}), as expected from the phase matching condition (\ref{eq:TAMM_DISP_graphene4}). 
The systematic deviation of the reflectivity dip position with respect to the OTS frequency predicted by the phase matching condition is due to the approximation made in writing Eq. (\ref {eq:TAMM_DISP_graphene4})  and the finite thickness of the GaAs "barrier". Interestingly, the effect of the graphene sheet on the OTS frequency depends upon its position inside the microcavity, being the largest when it is attached to the surface of the Bragg reflector  (Fig. \ref{fig:graphene_sheet2}(b)).
It happens because the phase shift introduced by graphene depends on the induced curremt, proportional to the transverse electric field, which is the largest near the BR surface.

%
%
{\it \textbf{Discussion.}} Our calculated results show that Tamm polaritons can be supported by structures where the EM field confinement outside the Bragg reflector is provided by the nearly full reflection that  
takes place in the reststrahlen band of a polar semiconductor.
Even though the GaAs slab considered here as an example is not a perfect mirror, the phase matching condition is a robust method for determining the OTS, as confirmed by the direct calculation of  the reflectivity of the whole structure (air-GaAs-cavity-BR-air). The spectrum of the latter shows a characteristic sharp dip related to the Tamm polariton. The quality factor of such a Tamm microcavity can be estimated by the ratio between the optical phonon damping parameter and typical reststrahlen band frequencies, $(1-Q) \sim \Gamma _{\text{TO}}/\omega_{\text{TO}}\approx 0.015$ for GaAs at room temperature.\\
\noindent
The microcavity mode spectral position (within the GaAs reststrahlen band) depends on the cavity width ($\delta=0$) and, additionally, can be adjusted by inserting a graphene layer and changing the graphene's Fermi energy, $E_F$. The control via tuning $E_F$ (usually achieved by varying the gate voltage applied to graphene \cite{neto_guinea_peres_novoselov_geim_2009}) is more efficient if the graphene sheet is placed onto the BR surface (Fig. \ref{fig:graphene_sheet2}(b)).  
This situation is also easier to implement in practice, since graphene can be simply transferred to the BR surface and then covered by an appropriate dielectric, e.g. a polymer layer.
The electrical tuning of the Fermi level can be realized in a back gate configuration by using a moderately doped GaAs, avoiding coupling of GaAs plasmons to the optical phonons. 
It is worth noting that the interaction of the Tamm polariton with graphene (Drude-type) plasmons introduces an additional phase shift and, consequently, shifts the cavity mode frequency. However, it does not produce graphene surface plasmon-polaritons (evanescent waves centred on the graphene sheet) unless the transverse wavevector, $q$, becomes larger than $\sqrt {\epsilon _{\text {m}}}\omega /c$ (where $\epsilon _{\text {m}}$ is the largest of the dielectric constants of the materials cladding the graphene sheet) \cite{bludov_2013}.\\
For $\delta=0$ the OTS dispersion, $\omega (q)$ is approximately parabolic with a positive effective mass, similar to that obtained for "conventional" Tamm polaritons, such as those occurring in a gold-GaAs/AlAs superlattice heterostructure \cite{TammPlasmonTP}. However, for $\delta \neq 0$ we find that deviations from this behavior become strong and a region of $q$ appears where the group velocity becomes negative (Fig. \ref{fig:disp_rel_delta}), i.e. the OTS becomes a backward wave. Interestingly, Tamm polariton modes with negative group velocity have been predicted for a terminated superlattice with the period composed of a graphene sheet and a non-dispersive dielectric layer \cite{Smirnova_2014}.\\
In conclusion, the performed calculations show that one can make a Tamm microcavity in the FIR region by using GaAs (or other polar semiconductor) as a phononic mirror in addition to a Bragg reflector. The insertion of a graphene sheet in the microcavity offers the advantage of cavity mode tunability. While Tamm states in the near-IR spectral range have been observed experimentally, we hope that this work will stimulate experimental studies in the  longer wavelength range. 
A cavity supporting such resonant modes can be useful for making a tunable light source in the spectral range where it is still a challenge.
\\

%
{\it \textbf{Acknowledgments.}}
Funding from the European Commission, within the project "Graphene-Driven Revolutions in ICT and Beyond" (Ref. No. 696656), and from the Portuguese Foundation for Science and
Technology (FCT) in the framework of the Strategic Funding UID/FIS/04650/2013 is gratefully acknowledged. The authors are grateful to Prof. Nuno Peres and Dr. Yuliy Bludov for helpful discussions. 
\\

{\it \textbf{Appendix A: Fresnel coefficients in case of oblique incidence.}} 
\setcounter{equation}{0}
When we consider oblique incidence we must take into account the contributions of the electric and magnetic fields perpendicular and parallel to the interface surface between the two dielectric mediums. Thus, we must consider two different polarizations: {\textit s}-polarization (or TE waves) and {\textit p}-polarization (or TM waves). 
An arbitrarily polarized field plane wave can be written as the superposition of two orthogonally polarized plane waves:
\begin{align}
	\mathbf{E}= \mathbf{E}^{(s)} + \mathbf{E}^{(p)},
\end{align}
where $\mathbf{E}^{(s)}$ and $\mathbf{E}^{(p)}$ represent the perpendicular and parallel electric fields, respectively, relative to the incidence plane interface.
The reflection and transmission Fresnel coefficients are defined as \cite{novotny_nanooptics}:
\begin{align}
	&\mathbf{E}_r = \hat{r}^{(s)} \cdot \mathbf{E}_i \label{eq:FCrs} \\ 
	&\mathbf{H}_r = \hat{r}^{(p)} \cdot \mathbf{H}_i \label{eq:FCrp} 
\end{align}
and
\begin{align}
	&\mathbf{E}_t = \hat{t}^{(s)} \cdot \mathbf{E}_i \label{eq:FCts}. \\
	&\mathbf{H}_t = \hat{t}^{(p)} \cdot \mathbf{H}_i \label{eq:FCtp}, 
\end{align} 
where $\hat{r}^{(s)}$ and $\hat{r}^{(p)}$ are the Fresnel reflection coefficients and $\hat{t}^{(s)}$ and $\hat{t}^{(p)}$ are the Fresnel transmission coefficients for $s$ and $p$-polarization, respectively.\\
The context of this work requires the study of {\textit p}-polarization. In this case the $\mathbf{H}$ field is parallel to the surface:
\begin{align}
	&\mathbf{H}_i^{(p)} = \hat{y} H_i e^{i \left( \mathbf{k}_i \cdot \mathbf{r} - \omega t \right)} \label{eq:ppol_M1} \\
	&\mathbf{H}_r^{(p)} = \hat{y} H_r e^{i \left( \mathbf{k}_r \cdot \mathbf{r} - \omega t \right)} \label{eq:ppol_M2} \\
	&\mathbf{H}_t^{(p)} = \hat{y} H_t e^{i \left( \mathbf{k}_t \cdot \mathbf{r} - \omega t \right)} \label{eq:ppol_M3} \text{ ,}
\end{align}
where the subscripts $i,r \text{ and } t$ correspond to the incident, reflected and transmitted fields, respectively.
\\

\begin{figure}[t]
	\centering
	\includegraphics[scale=0.45]{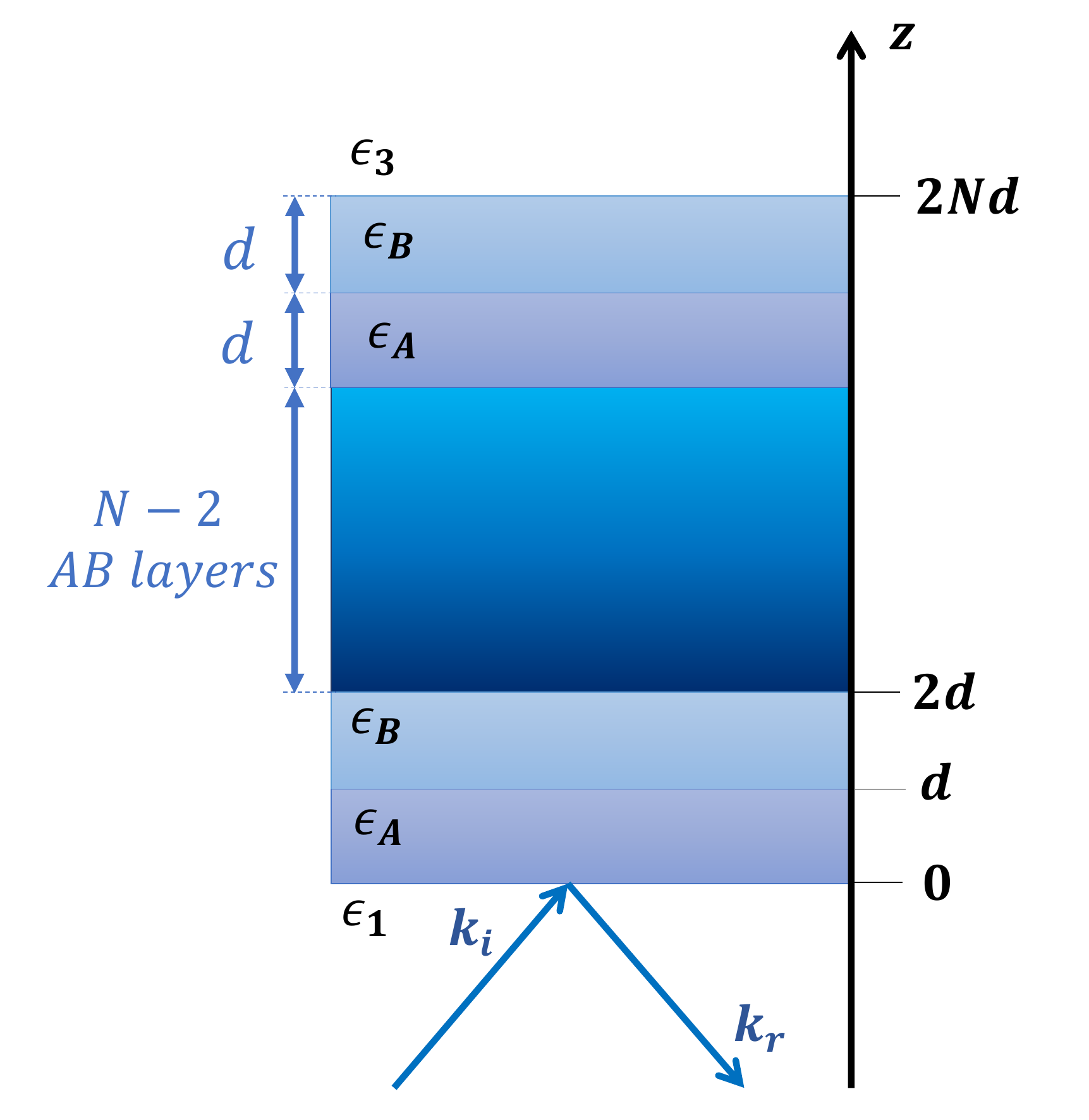}
	\caption{Scheme corresponding to a $N$-bilayer Bragg reflector.}
	\label{fig:bragg_ref}
\end{figure}

\begin{figure}[t]
	\centering
	\includegraphics[scale=0.6]{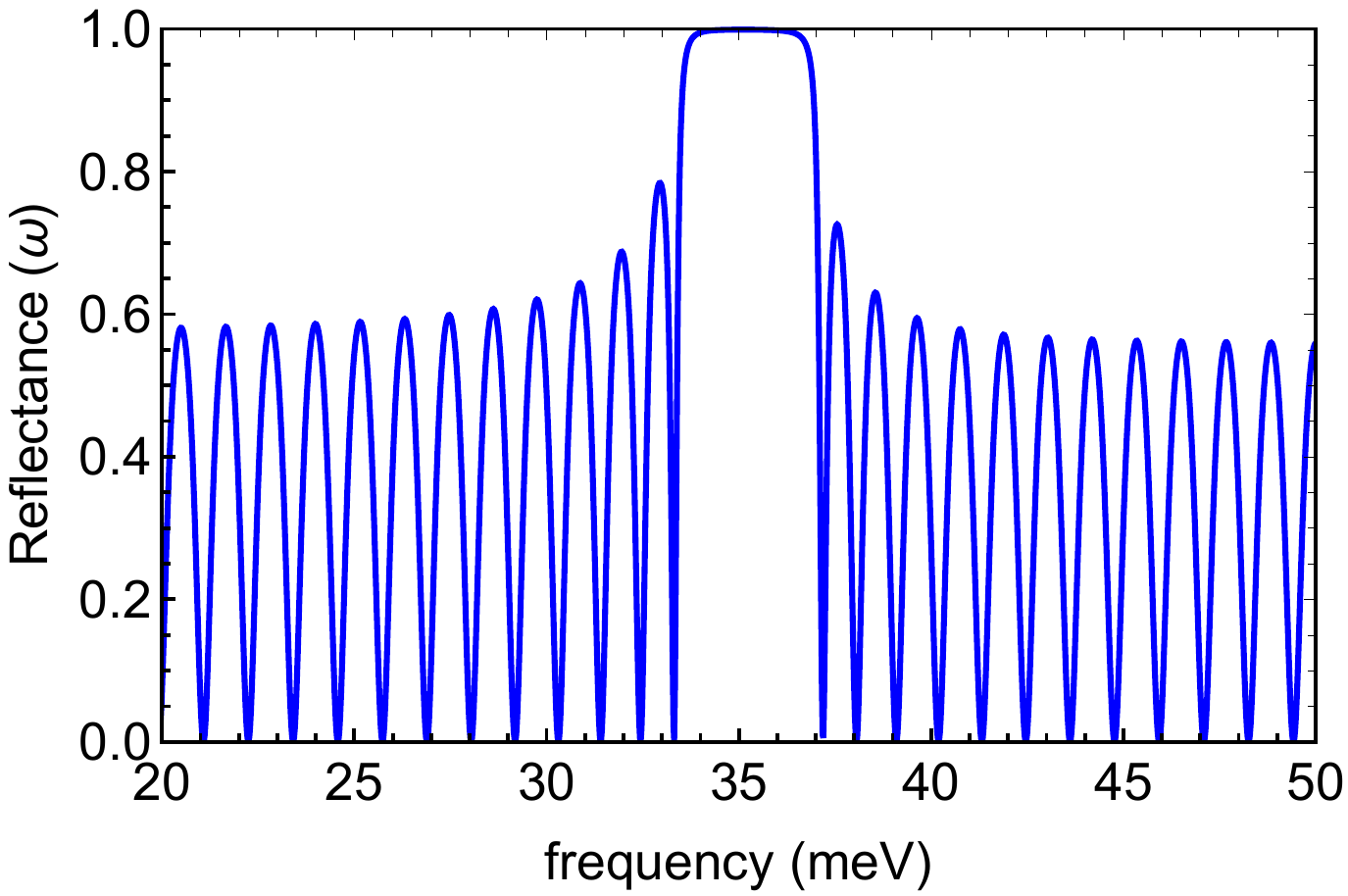}
	\caption{Reflectivity spectrum of a Si/Ge BR with $d=2.4\mu$m and $N$=30.}
	\label{fig:BR_reflect}
\end{figure}

{\it \textbf{Appendix B: Transfer matrix method for a Bragg reflector.}}
Here we are going to derivate the global transfer matrix of a Bragg reflector by relating the electromagnetic fields at the surface of the Bragg reflector with the electromagnetic fields located in the last layer of a finite $N$-period Bragg reflector. This has been done previously\cite{icton2018}.\\ 
The transfer matrix of a one period Si/Ge superlattice can be determined by the following boundary conditions:\\
\[ 
\left(
\begin{tabular}{ccc}
$H_1 (\mathbf{r},t)$  \\
$E_{1x}(\mathbf{r},t)$   
\end{tabular}
\right)_{z=0^{-}}
=
\left(
\begin{tabular}{ccc}
$H_{A} (\mathbf{r},t)$  \\
$E_{Ax}(\mathbf{r},t)$  
\end{tabular}
\right)_{z={0^{+}}}\text{,} \addtag \label{TM_method1}
\] \\
where $H_1 (\mathbf{r},t)$ and $E_{1x} (\mathbf{r},t)$ correspond to the fields propagating in air (z$<$0), and $H_{A} (\mathbf{r},t$ and $E_{Ax}(\mathbf{r},t)$ to the fields propagating in medium A. Furthermore,
\[
\left(
\begin{tabular}{ccc}
$H_{A} (\mathbf{r},t)$  \\
$E_{Ax}(\mathbf{r},t)$  
\end{tabular}
\right)_{z={0^{+}}} 
=
\hat{T}_{A}^{-1} \cdot
\left(
\begin{tabular}{ccc}
$H_{A} (\mathbf{r},t)$  \\
$E_{Ax}(\mathbf{r},t)$  
\end{tabular}
\right)_{z={d^{-}}} \text{,} \addtag \label{TM_method2} 
\]
\[
\hat{T}_{A}^{-1} \cdot
\left(
\begin{tabular}{ccc}
$H_{A} (\mathbf{r},t)$  \\
$E_{Ax}(\mathbf{r},t)$  
\end{tabular}
\right)_{z={d^{-}}}
=
\hat{T}_{A}^{-1}
\left(
\begin{tabular}{ccc}
$H_{B} (\mathbf{r},t)$  \\
$E_{Bx}(\mathbf{r},t)$  
\end{tabular}
\right)_{z={d^{+}}} \text{,} \addtag \label{TM_method3} 
\] \\ 
and
\[
\left(
\begin{tabular}{ccc}
$H_{B} (\mathbf{r},t)$  \\
$E_{Bx}(\mathbf{r},t)$  
\end{tabular}
\right)_{z={d^{+}}}
=
\hat{T}_{B}^{-1} \cdot
\left(
\begin{tabular}{ccc}
$H_{B} (\mathbf{r},t)$  \\
$E_{Bx}(\mathbf{r},t)$  
\end{tabular}
\right)_{z={2d^{-}}} \text{,}  \addtag \label{TM_method4}
\] \\
where $H_{B} (\mathbf{r},t)$ and $E_{Bx}(\mathbf{r},t)$ correspond to the fields propagating in medium B.   
By analyzing expressions (\ref{TM_method1}), (\ref{TM_method2}), (\ref{TM_method3}) and (\ref{TM_method4}) one obtains:
\[ 
\left(
\begin{tabular}{ccc}
$H_1 (\mathbf{r},t)$  \\
$E_{1x}(\mathbf{r},t)$   
\end{tabular}
\right)_{z=0^{-}}
= 
\hat{T}_{A}^{-1}\cdot \hat{T}_B^{-1} \cdot
\left(
\begin{tabular}{ccc}
$H_{B} (\mathbf{r},t)$  \\
$E_{Bx}(\mathbf{r},t)$  
\end{tabular}
\right)_{z={2d^{-}}} \text{,} \addtag \label{TM_method5}
\]
where the matrices $\hat{T}_{A}^{-1}$ and $\hat{T}_{B}^{-1}$ are given by: \par
\[ 
\hat{T}_{A,B}^{-1} = 
\left(
\begin{tabular}{ccc}
$\cos\left( k_{A,B}d \right)$ & $ i\frac{\omega}{c} \frac{\epsilon_{A,B} }{k_{A,B}} \sin(k_{A,B} d)$   \\
$i \frac{c}{\omega} \frac{k_{A,B}}{\epsilon_{A,B} } \sin(k_{A,B} d)$ & $\cos\left( k_{A,B} d \right)$
\end{tabular}
\right) \text{.} \addtag \label{TM_AB}
\] 
%
%
with 
\begin{align}
	k_A=\sqrt{\epsilon_A^2  \left( \frac{\omega}{c} \right) ^2 - q^2 \left( \theta, \omega \right) } \text{ ,} \\
	k_B=\sqrt{\epsilon_B^2  \left( \frac{\omega}{c}\right)^2 - q^2 \left( \theta, \omega \right) } \text{ .} 
\end{align}
In case of a $N$-period lattice, figure \ref{fig:bragg_ref}, 
one can generalize equation (\ref{TM_method5}) for a \textit{N}-period structure:
\[ 
\left(
\begin{tabular}{ccc}
$H_1 (\mathbf{r},t)$  \\
$E_{1x}(\mathbf{r},t)$   
\end{tabular}
\right)_{z=0^{-}}
=
\left( \hat{T}_{A}^{-1} \cdot \hat{T}_B^{-1} \right)^N \cdot
\left(
\begin{tabular}{ccc}
$H_{B} (\mathbf{r},t)$  \\
$E_{Bx}(\mathbf{r},t)$  
\end{tabular}
\right)_{z={2Nd^{-}}} \text{.} \addtag \label{TM_method1N}
\]
The cladding media fields can be matched in the following way:
\[ 
\left(
\begin{tabular}{ccc}
$H_1 (\mathbf{r},t)$  \\
$E_{1x}(\mathbf{r},t)$   
\end{tabular}
\right)_{z=0^{-}}
=
\left( \hat{T}_{A}^{-1}\cdot \hat{T}_B^{-1} \right)^N \cdot
\left(
\begin{tabular}{ccc}
$H_{3} (\mathbf{r},t)$  \\
$E_{3x}(\mathbf{r},t)$  
\end{tabular}
\right)_{z={2Nd^{+}}}\text{,} \addtag \label{TM_method2N}
\]
which translates to
\[ 
\left(
\begin{tabular}{ccc}
$1 + \hat{r}_p$  \\
$\frac{c k_{1z}}{\epsilon_1 \omega}(1-\hat{r}_p)$   
\end{tabular}
\right)
=
\hat{T}_N^{-1} \cdot
\left(
\begin{tabular}{ccc}
$\hat{t}_p$  \\
$- \frac{c k_{3z}}{\epsilon_3 \omega}\hat{t}_p$   
\end{tabular}
\right) \text{,}  \addtag \label{TM_method3N}
\]
where $\hat{T}_N^{-1}$ is the inverse tranfer matrix of the whole structure:
\begin{align}
	\hat{T}_N^{-1} \equiv \left( \hat{T}_{A}^{-1}\cdot \hat{T}_B^{-1} \right)^N \text{.}
	\label{eq:TM_N}
\end{align}
From equation (\ref{TM_method3N}) one readily obtains the Fresnel coeficients of the structure. The reflectivity $\left( \text{R}=\vert \hat{r}^{(p)} \vert ^2 \right)$ spectrum of a DBR is shown in figure \ref{fig:BR_reflect}.
\\

\begin{figure}[t]
	\centering
	\includegraphics[scale=0.4]{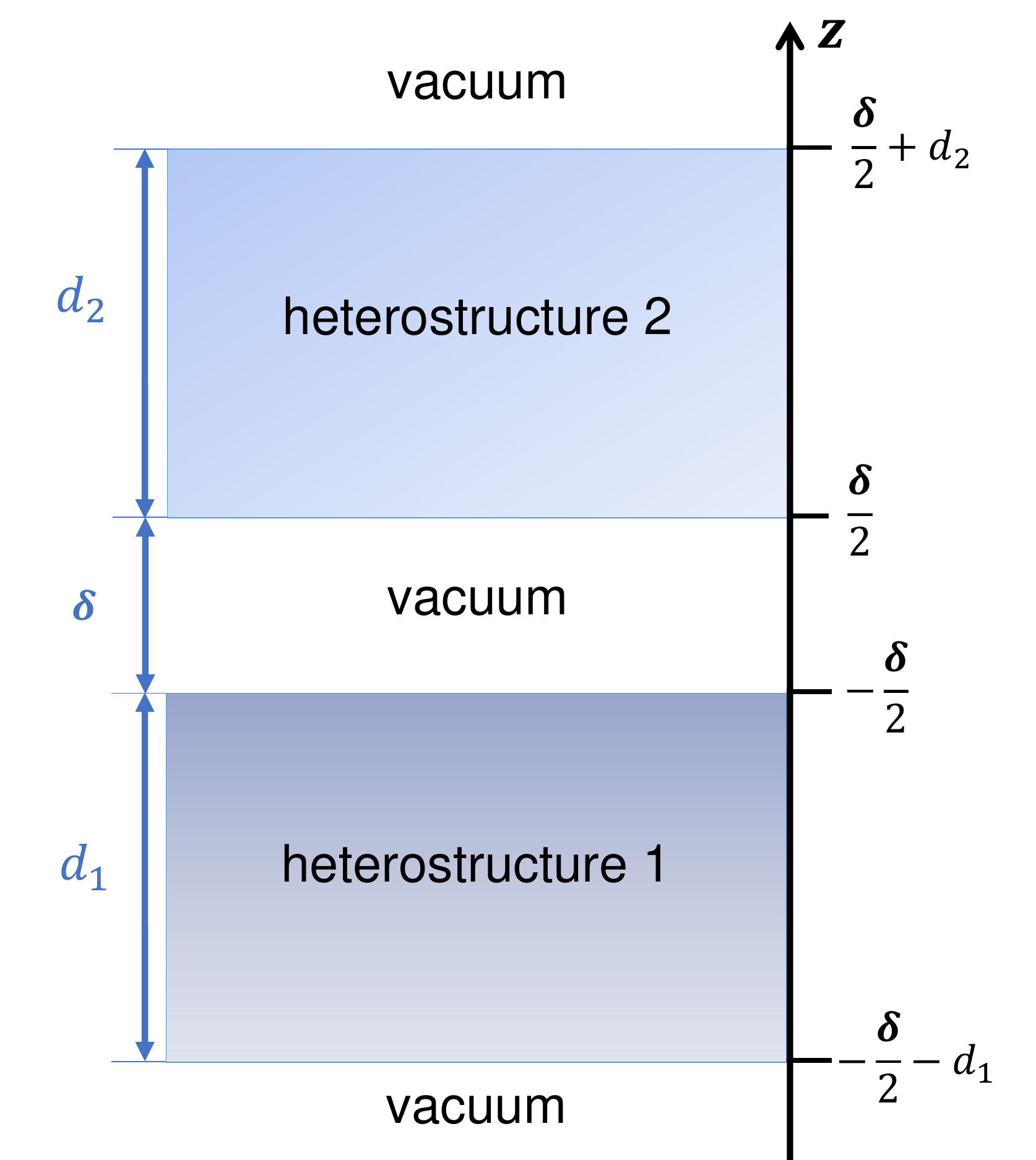}
	\caption{Scheme corresponding to a structure constituted by two different heterostructures separated by a gap.}
	\label{fig:disp_rel_ref}
\end{figure}

{\it \textbf{Appendix C: Derivation of the phase matching relation.}} 
Here we present the derivation of the phase matching equation. Using the transfer matrix method one has:
\[ 
\left(
\begin{tabular}{ccc}
$H_y $  \\
$E_x$   
\end{tabular}
\right)_{z=-\frac{\delta }{2}}
= \hat{T}_1^{-1} \cdot
\left(
\begin{tabular}{ccc}
$H_y $  \\
$E_x $  
\end{tabular}
\right)_{z=-d_1-\frac{\delta }{2}} \text{,} \addtag \label{eq:tamm_pol_bc1}
\]
which gives us
\[ 
\left(
\begin{tabular}{ccc}
$H_y $  \\
$E_x$   
\end{tabular}
\right)_{z=-\frac{\delta }{2}}
= \hat{T}_1^{-1} \cdot
\left(
\begin{tabular}{ccc}
$\hat{t}^{(p)}_1$  \\
$- \hat{t}^{(p)}_1 \frac{ck}{\omega}$  
\end{tabular}
\right) \text{ ,} \addtag \label{eq:tamm_pol_t1} 
\]
where $\hat{t}^{(p)}_1$ is the complex amplitude of the magnetic field of the wave outgoing from the lower surface of the heterostructure 1, figure \ref{fig:disp_rel_ref}. 
%
Similarly, we have:
\[ 
\left(
\begin{tabular}{ccc}
$H_y $  \\
$E_x$   
\end{tabular}
\right)_{z=\frac{\delta }{2}}
= \hat{T}_2 \cdot
\left(
\begin{tabular}{ccc}
$H_y $  \\
$E_x $  
\end{tabular}
\right)_{z=d_2+\frac{\delta }{2}} \text{,} \addtag \label{eq:tamm_pol_bc2}
\]
i.e.
\[ 
\left(
\begin{tabular}{ccc}
$H_y $  \\
$E_x$   
\end{tabular}
\right)_{z=\frac{\delta }{2}}
= \hat{T}_2 \cdot
\left(
\begin{tabular}{ccc}
$\hat{t}^{(p)}_2$  \\
$\hat{t}^{(p)}_2 \frac{ck}{\omega}$  
\end{tabular}
\right) \text{ ,} \addtag \label{eq:tamm_pol_t2}
\]
where $\hat{t}^{(p)}_2$ is the amplitude of the wave outgoing from the upper surface (it propagates along the positive direction of the $z-$ axis). Notice that there is no incoming wave in this case. \par
The fields at $z=\pm \delta /2$ are related by
\[ 
\left(
\begin{tabular}{ccc}
$H_y $  \\
$E_x$   
\end{tabular}
\right)_{z=\frac{\delta }{2}}
=
\left(
\begin{tabular}{ccc}
$\cos (k \delta)$ & $i \frac{\omega }{c k} \sin(k \delta )$  \\
$i \frac{ck}{\omega} \sin (k \delta)$ & $\cos (k \delta )$
\end{tabular}
\right) \cdot
\left(  
\begin{tabular}{ccc}
$H_y$ \\
$E_x$
\end{tabular}
\right)_{z=-\frac{\delta }{2}} \text{ ,} \addtag \label{eq:TM_delta1}
\]
\[
\left(
\begin{tabular}{ccc}
\resizebox{0.05\textwidth}{!}{$\cos (k \delta)$} & \resizebox{0.07\textwidth}{!}{$i \frac{\omega }{c k} \sin(k \delta )$}  \\
\resizebox{0.07\textwidth}{!}{$i \frac{ck}{\omega} \sin (k \delta)$} & \resizebox{0.05\textwidth}{!}{$\cos (k \delta )$}
\end{tabular}
\right) \cdot
\left(  
\begin{tabular}{ccc}
$H_y$ \\
$E_x$
\end{tabular}
\right)_{z=-\frac{\delta }{2}}
\equiv
\hat{T}_{\delta }^{-1} \cdot
\left(
\begin{tabular}{ccc}
$H_y$ \\
$E_x$
\end{tabular}
\right)_{z=-\frac{\delta }{2}} \text{.} \addtag \label{eq:TM_delta2}
\]
Combining (\ref{eq:tamm_pol_t1}), (\ref{eq:tamm_pol_t2}), (\ref{eq:TM_delta1}) and (\ref{eq:TM_delta2}), we have:
\[ 
\hat{T}_2 \cdot
\left(
\begin{tabular}{ccc}
$\hat{t}^{(p)}_2$  \\
$\hat{t}^{(p)}_2 \frac{c k}{\omega}$   
\end{tabular}
\right)
= \hat{T}_{\delta }^{-1} \cdot \hat{T}_1^{-1} \cdot
\left(
\begin{tabular}{ccc}
$\hat{t}^{(p)}_1$  \\
$- \hat{t}^{(p)}_1 \frac{ck}{\omega}$  
\end{tabular}
\right) \text{.} \addtag \label{eq:tamm_pol_t3}
\]
By setting $ \hat{t}^{(p)}_1 = \text{A } \hat{t}^{(p)}_2$, where A is a complex number, we obtain:
\[
\hat{T}_1 \cdot \hat{T}_{\delta } \cdot  \hat{T}_2 \cdot
\left(
\begin{tabular}{ccc}
$1$\\
$\frac{c k}{\omega}$
\end{tabular}
\right)
=
A
\left(
\begin{tabular}{ccc}
$1$\\
$-\frac{c k }{\omega}$
\end{tabular}
\right) \text{,}\addtag \label{eq:tamm_pol_A}
\]
or explicitly
\begin{equation}
\label{eq:tamm_A_amplitude}
\begin{cases}
(\hat{T}_G)_{11} + (\hat{T}_G)_{12} \frac{c k}{\omega} = A  \text{ ,}\\
\frac{\omega }{c k} (\hat{T}_G)_{21} + (\hat{T}_G)_{22}=-A \text{ ,}
\end{cases}
\end{equation}
where
\begin{equation}
\label{eq:tamm_pol_transfermatrix}
\hat{T}_G \equiv \hat{T}_1 \hat{T}_{\delta } \hat{T}_2 \text{.}
\end{equation}
The only variable in equations (\ref{eq:tamm_A_amplitude}) is $A$, so they are compatible if
\begin{equation}
\label{tamm_states_cond}
(\hat{T}_G)_{11}+\frac{c k}{\omega} (\hat{T}_G)_{12} + \frac{\omega }{c k} (\hat{T}_G)_{21}+(\hat{T}_G)_{22}=0 \text{ .}
\end{equation}
Since $k=\sqrt{(\omega / c )^2 - q^2}$, this is an implicit dispersion relation, $\omega (q)$, of the Tamm mode. We recognize the left-hand side of (\ref{tamm_states_cond}) as the denominator of the Fresnel transmission coefficient of the whole structure. \par
The dispersion relation can be cast in a more convenient form:
\[
\hat{T}^{-1} \cdot
\left(
\begin{tabular}{ccc}
$\hat{t}^{(p)}$\\
$-\frac{c k}{\omega} \hat{t}^{(p)}$
\end{tabular}
\right)
=
\left(
\begin{tabular}{ccc}
$1 + \hat{r}^{(p)}$\\
$\frac{c k }{\omega} (\hat{r}^{(p)}-1)$
\end{tabular}
\right) \text{.}\addtag \label{eq:tamm_pol1}
\]
Similarly,
\[
\hat{T} \cdot
\left(
\begin{tabular}{ccc}
$\hat{t}^{(p)}$\\
$\frac{c k}{\omega} \hat{t}^{(p)}$
\end{tabular}
\right)
=
\left(
\begin{tabular}{ccc}
$1 + \hat{r}^{(p)}$\\
$\frac{c k }{\omega} (1-\hat{r}^{(p)})$
\end{tabular}
\right) \text{.}\addtag \label{eq:tamm_pol2}
\]
Using relations (\ref{eq:tamm_pol1}) and (\ref{eq:tamm_pol2}), we can rewrite (\ref{eq:tamm_pol_t3}) in terms of the amplitudes of the reflected waves:
\[
\hat{T}_{\delta } \cdot
\left(
\begin{tabular}{ccc}
$1+\hat{r}^{(p)}_2$\\
$\frac{c k}{\omega}(1- \hat{r}^{(p)}_2 )$
\end{tabular}
\right)
=
B
\left(
\begin{tabular}{ccc}
$1 + \hat{r}^{(p)}_1$\\
$\frac{c k }{\omega} (\hat{r}^{(p)}_1 -1)$
\end{tabular}
\right) \text{.}\addtag \label{eq:tamm_pol3}
\]
or, explicitly,
\begin{equation}
\begin{cases}
\resizebox{0.4\textwidth}{!}{$\cos (k \delta ) (1+\hat{r}^{(p)}_2) - i \sin (k \delta) (1-\hat{r}^{(p)}_2) = B (1+\hat{r}^{(p)}_1) \text{ ,}$} \\
\resizebox{0.4\textwidth}{!}{$-i \sin (k \delta) (1+\hat{r}^{(p)}_2) + \cos(k \delta) (1- \hat{r}^{(p)}_2)=B (\hat{r}^{(p)}_1 -1) \text{ ,}$}
\end{cases}
\end{equation}
which translates to
\begin{equation}
\label{eq:tamm_B_amplitude_solved}
\begin{cases}
e^{-ik \delta } + \hat{r}^{(p)}_2 e^{i k \delta} = B (1+\hat{r}^{(p)}_1) \text{,} \\
e^{-ik \delta } - \hat{r}^{(p)}_2 e^{i k \delta} =B (\hat{r}^{(p)}_1 -1) \text{,}
\end{cases}
\end{equation}
where $B$ is a constant.
The condition of compatibility of equations (\ref{eq:tamm_B_amplitude_solved}) reads:
\begin{equation}
\label{eq:TAMM_DISP}
\hat{r}^{(p)}_1 \hat{r}^{(p)}_2 e^{2 i k \delta }=1 \text{ .}
\end{equation}
\qquad

{\it \textbf{Appendix D: Derivation of the phase matching condition when introducing a graphene sheet.}}
In the case of having a graphene sheet located in the center of the microcavity, one can derive the dispersion relation from the following equation:
\[
\hat{T}_{\frac{\delta}{2}} \cdot \hat{T}_{Graph} \cdot \hat{T}_{\frac{\delta}{2}} \cdot
\left(
\begin{tabular}{ccc}
$1+\hat{r}^{(p)}_2$\\
$\frac{c k}{\omega}(1- \hat{r}^{(p)}_2 )$
\end{tabular}
\right)
=
D
\left(
\begin{tabular}{ccc}
$1 + \hat{r}^{(p)}_1$\\
$\frac{c k }{\omega} (\hat{r}^{(p)}_1 -1)$
\end{tabular}
\right) \text{,}\addtag \label{eq:tamm_pol_graphene1}
\] 
where $D$ is a constant and $\hat{T}_{Graph}$ is the graphene's transfer matrix:
\[ 
\hat{T}_{Graph}
=
\left(
\begin{tabular}{ccc}
1 & $\frac{4 \pi}{c} \sigma \left( \omega \right)$  \\
0 & 1
\end{tabular}
\right) \text{ ,} \addtag \label{eq:TM_graphene}
\]
with $\sigma \left( \omega \right)$ being the graphene's 2D optical conductivity \cite{neto_guinea_peres_novoselov_geim_2009,spps_graphene}: 
\begin{equation}
\sigma \left( \omega \right) =  \sigma_0 \frac{4 E_f}{\pi} \frac{1}{\Gamma - i \hbar \omega} \text{ ,}
\end{equation}
where $\sigma_0=\pi e^2 / \left( 2 h \right)$, $e<0$ is the electron charge, $E_f>0$ denotes graphene's Fermi energy and $\Gamma$ corresponds to Drude's damping term.
It is more conveninent if we diagonalize $\hat{T}_{\frac{\delta}{2}}$:
\[
\hat{D}_{\frac{\delta}{2}}
=
\hat{M}^{-1} \cdot \hat{T}_{\frac{\delta}{2}} \cdot \hat{M}
=
\left(
\begin{tabular}{ccc}
$e^{i k \frac{\delta}{2}}$ & $0$  \\
$0$ & $e^{-i k \frac{\delta}{2}}$
\end{tabular}
\right) \text{,}\addtag \label{eq:tamm_pol_graphene2}
\] 
where $\hat{M}$ and $\hat{M}^{-1}$ are the rotational matrix and its inverse matrix, respectively:
\[
\hat{M}
=
\left(
\begin{tabular}{ccc}
$1$ & $1$  \\
$-\frac{ck}{\omega}$ & $\frac{ck}{\omega}$
\end{tabular}
\right) \text{,}\addtag \label{eq:tamm_pol_graphene3}
\]
and
\[
\hat{M}^{-1}
=
\left(
\begin{tabular}{ccc}
$\frac{1}{2}$ & $-\frac{\omega}{2 ck}$  \\
$\frac{1}{2}$ & $\frac{\omega}{2 ck}$
\end{tabular}
\right) \text{.}\addtag \label{eq:tamm_pol_graphene4}
\]
Knowing that
\begin{equation}
\hat{T}_{\frac{\delta}{2}} \cdot \hat{T}_{Graph} \cdot \hat{T}_{\frac{\delta}{2}} = \hat{M} \cdot \left( \hat{D}_{\frac{\delta}{2}} \cdot \hat{M}^{-1} \cdot \hat{T}_{Graph} \cdot \hat{M} \cdot \hat{D}_{\frac{\delta}{2}} \right) \hat{M}^{-1} \text{ ,}
\end{equation}
one can easily solve equation (\ref{eq:tamm_pol_graphene1}):
\begin{equation}
\label{eq:tamm_pol_graphene5}
\begin{cases}
\hat{r}^{(p)}_2 e^{ik\delta}\left(1- \frac{2 \pi k \sigma \left( \omega \right)}{\omega} \right)
&+ e^{-ik\delta} \left( \frac{2 \pi k \sigma \left( \omega \right)}{\omega} +1\right)  \\ 
&+ \frac{2 \pi \sigma \left( \omega \right) k}{\omega} - \hat{r}^{(p)}_2 \frac{2 \pi \sigma \left( \omega \right)k}{\omega}= \resizebox{0.08\textwidth}{!}{$D (1+\hat{r}^{(p)}_1)$} \text{ ,} \\
\\
\hat{r}^{(p)}_2 e^{ik\delta}\left(1- \frac{2 \pi k \sigma \left( \omega \right)}{\omega} \right)
&- e^{-ik\delta} \left( \frac{2 \pi k \sigma \left( \omega \right)}{\omega} +1\right) \\
&- \frac{2 \pi \sigma \left( \omega \right) k}{\omega} - \hat{r}^{(p)}_2 \frac{2 \pi \sigma \left( \omega \right)k}{\omega}=
\resizebox{0.08\textwidth}{!}{$ D (\hat{r}^{(p)}_1-1)$} \text{ ,} \nonumber
\end{cases}
\end{equation}
%
%
which translates to
\begin{equation}
\label{eq:TAMM_DISP_graphene1}
\resizebox{0.43\textwidth}{!}{$\hat{r}^{(p)}_1 \hat{r}^{(p)}_2 e^{i k \delta } \left(\frac{2 \pi \sigma \left( \omega \right) k}{\omega} -1 \right) - \frac{2 \pi \sigma \left( \omega \right) k}{\omega } \left(\hat{r}^{(p)}_1 + \hat{r}^{(p)}_2 \right) + e^{-ik\delta}\left( \frac{2 \pi \sigma \left( \omega \right) k}{\omega} + 1 \right) = 0 \text{ .}$}
\end{equation}
\vspace{1mm}

%
\bibliographystyle{unsrt}
\bibliography{ref}
\end{document}